\documentclass[prl,aps,final,twocolumn,showkeys,showpacs]{revtex4-1}

\usepackage{graphicx}

\begin{document}

\title{Interferometry with Two Pairs of Spin Correlated Photons}
\author{Mladen Pavi\v ci\'c}
\homepage{http://m3k.grad.hr/pavicic}
\affiliation{Institut f\"ur Theoretische Physik, 
TU Berlin, Hardenbergstra\ss e 36, D--10623 Berlin 12, Germany\\
Atominstitut der \"Osterreichischen Universit\"aten, 
Sch\"uttelstra\ss e 115, A--1020 Wien, Austria\\
and Chair of Physics, Faculty of Civil engineering, University of Zagreb, 
Zagreb, Croatia}
\author{Johann Summhammer}
\affiliation{Atominstitut der \"Osterreichischen Universit\"aten, 
Sch\"uttelstra\ss e 115, A--1020 Wien, Austria}

\begin{abstract}
We propose a new experiment employing two independent 
sources of spin correlated photon pairs. Two photons from different
unpolarized sources each pass through a polarizer to a detector. 
Although their trajectories never mix or cross they exhibit 
4th--order--interference--like correlations when the other two 
photons interfere on a beam splitter even when the latter two do 
not pass any polarizers at all. A wave packet calculation shows that 
the experiment permits a very discriminatory test of hidden variable 
theories. 
\end{abstract}


\maketitle

Higher order interference effects have been much investigated 
because their nonlocal nature provides a powerful tool
for testing hidden variables--theories 
\cite{ou88,man83,paul86,rari90,ozwm90,owzm90,om88,ohm88,ohm87,%
cst90,hsz89,wa91,hv91,wzm91,silv93}. In a recent test 
Ou and Mandel \cite{om88} used the polarization correlation of
signal and idler photon of downconverted light and found a violation
of Bell's inequality by about six standard deviations. However their 
detectors were not sufficiently efficient. The idea
was extended to two--particle interferometry in a proposal 
by Horne, Shimony and Zeilinger \cite{hsz89}. In another recent 
experiment Wang, Zou, and Mandel \cite{wa91} tested the so-called 
de Broglie--Bohm pilot wave theory. The result was negative but the 
set--up was recognized to lack generality by Holland and Vigier 
\cite{hv91} and by Wang, Zou, and Mandel \cite{wzm91}. Thus 
excluding realistic nonlocality remains a challenge \cite{bell86}.  

In this paper we propose an experiment which should be the 
first realization of the 4th order interference of randomly prepared 
independent photons correlated in polarization and coming from 
independent sources. The experiment is based on a newly 
discovered interference effect of the 4th order on a beam 
splitter \cite{p-pra94}. The essential new element of the experiment
is that it puts together systems which were not in any way influenced 
by preparation, makes them interact, and then allows us inferring 
overall polarization (spin) correlations --- from the assumed 
quantum mechanical description of the unknown initial states --- by 
simultaneous measurement of four photons separated in space. 
Particular polarization (spin) correlations, 
unexpectedly found between photons which did not in any way directly 
interact and on the distant pairs of which polarization has not been 
measured at all, are expected to be confirmed by a future experiment.
Such experiments might eventually disprove any realistic hidden 
variable theory. 

\begin{figure}[htp]
\begin{center}
{\includegraphics[width=0.48\textwidth,height=0.26\textwidth]{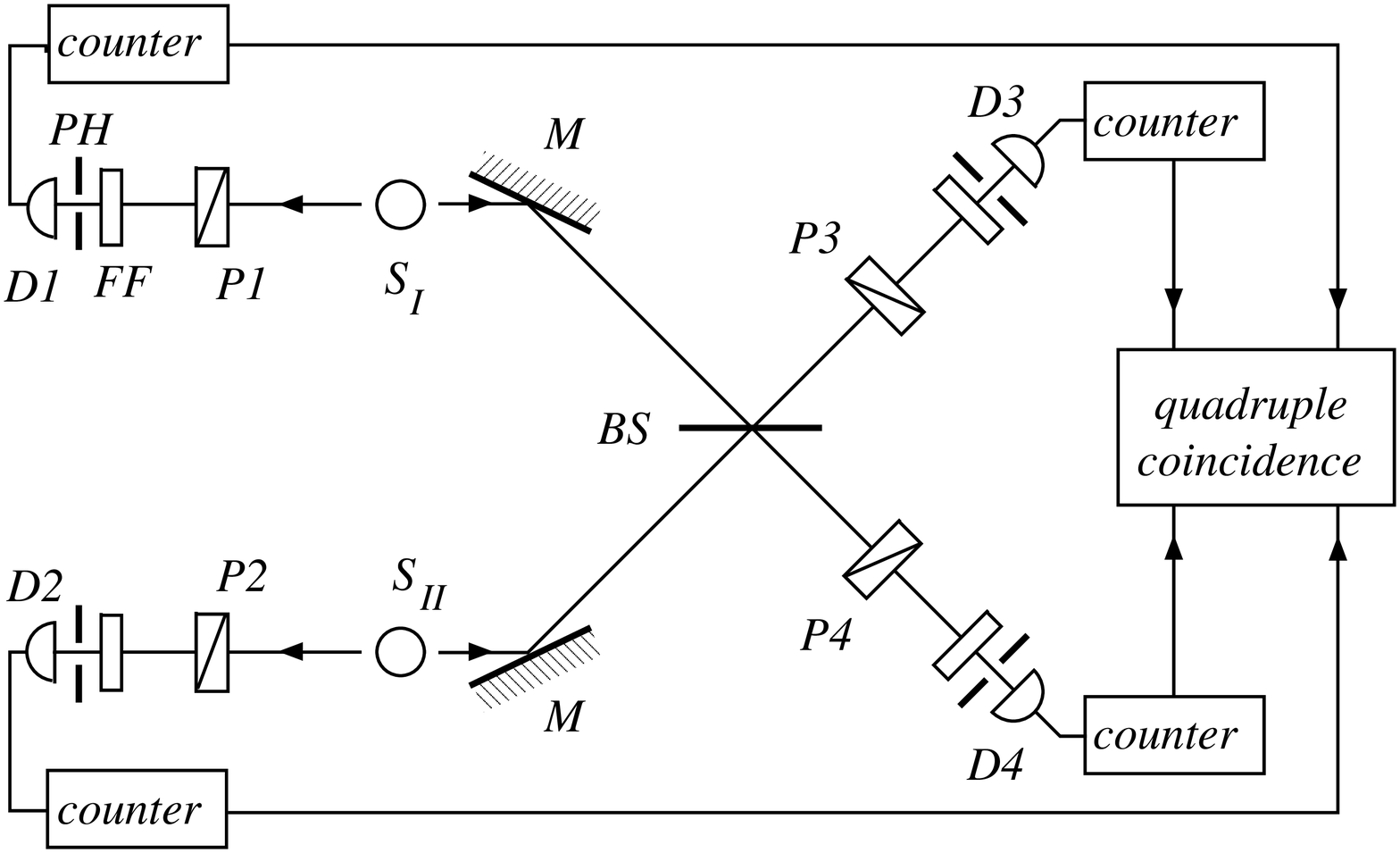}}
\end{center}
\caption{Lay--out of the proposed experiment.}
\label{exp}
\end{figure}

A schematic representation of the experiment is shown in 
Fig.~\ref{exp}. Two independent sources, $S_I$ and $S_{II}$, both 
simultaneously emit two photons correlated in polarization to the 
left and right. On the left photons we measure polarizations by 
the polarization filters P1 and P2 and on the right photons by P3 
and P4. Because of the beam splitter the paths leading through P3 
and P4 are available to the right photons from both source $S_{I}$ 
and source $S_{II}$. The resulting 4th order interference will 
manifest itself in the probability of quadruple coincidence counts 
in detectors D1, D2, D3 and D4. 

The sources can be atoms exhibiting cascade emission.
(Downconversion is not possible because it gives polarized photons.) 
The atoms of the two sources could be pumped to an upper level by two 
independent lasers \cite{ohm87}. This level would decay by emitting 
two photons correlated in polarization \cite{aspect2}. The independence 
of the two sources can be assured by slight differences in central 
frequency and drift of the two pump lasers. Hence there should
be no 2nd order interference at detectors D3 and D4, which could also 
be suppressed when the size of the sources exceeds the coherence 
length of the emitted photons. Interference of 4th
order would still occur, because its relevant coherence time 
is given by the inverse of the frequency difference of the two 
correlated photons \cite{ou88,man83,paul86,rari90}. 

The state of the four photons immediately after leaving the sources 
is described by the product of two entangled states:
\begin{eqnarray}
|\Psi\rangle&=&{1\over\sqrt2}\left(|1_x\rangle_1|1_x\rangle_3\>+
\>\,|1_y\rangle_1|1_y\rangle_3\right)\nonumber\\
&&\otimes{1\over\sqrt2}\left(|1_x\rangle_2|1_x\rangle_4\>+
\>\,|1_y\rangle_2|1_y\rangle_4\right)\label{eq:1}
\end{eqnarray}
Here, $|_x\rangle$ and $|_y\rangle$ 
denote the mutually orthogonal photon states. So, e.g., 
$|1_x\rangle_1$ means the state of 
photon 1 leaving the source $S_I$ to the left polarized in  
direction~$x$. In the following we use the annihilation 
operator formalism, often employed in quantum optical 
analysis, e.g.~by Paul \cite{paul86}, 
Mandel's group \cite{ozwm90,owzm90,ohm87}, 
and Campos et al. \cite{cst90}. 
The operator describing the polarization at P1 oriented along the 
$x$-axis and the subsequent detection at D1 acts as follows: 
${\hat a}_{1x}|1_{x}\rangle_1=
|0_{x}\rangle_1$, \ 
${\hat a}_{1x}^{\dagger}|0_{x}\rangle_1=
|1_{x}\rangle_1$, \  
${\hat a}_{1x}|0_{x}\rangle_1=0$, etc.~\cite{paul86}.   
When P1 is oriented at some angle $\theta_1$ polarization and 
detection are represented by 
${\hat a}_{1}={\hat a}_{1x}\cos\theta_1+{\hat a}_{1y}\sin\theta_1$. 
The phase the photon accumulates between the 
source $S_I$ and the detector D1 adds the factor 
$e^{i\omega_1(r_1/c+t^I_0-t_1)}$, where $\omega_1$ is the frequency 
of photon 1, $r_1$ is the path length from $S_I$ to D1, $c$ is the 
velocity of light, $t^I_0$ is the time of emission of a pair of 
photons at $S_I$, and $t_1$ is the time of detection at D1. Hence 
the annihilation of a photon at detector D1 means application of the 
operator 
${\hat E}_1=({\hat a}_{1x}\cos\theta_1+{\hat a}_{1y}\sin\theta1)
e^{i\omega_1(r_1/c+t^I_0-t_1)}$ onto the initial state of 
Eq.~(\ref{eq:1}). 
Similarly, detection of photon 2 at D2 means application of 
${\hat E}_2=({\hat a}_{2x}\cos\theta_2+{\hat a}_{2y}\sin\theta_2)
e^{i\omega_2(r_2/c+t^{II}_0-t_2)}$, where the symbols are defined by 
analogy. On the right side of the sources, a detection at D3 can be 
caused by photon 3 emitted by source $S_I$ or by photon 4 
emitted by source $S_{II}$. The beamsplitter BS may have polarization 
dependent transmission and reflection coefficients, denoted by $T_x$, 
$T_y$, and $R_x$, $R_y$, respectively. The angle of the polarizer P3 
is given by $\theta_3$. Hence we obtain \footnote{For the given operators 
                    cf.~Refs.~\protect\cite{ou88,owzm90,om88,ohm88} 
                    Note that positive imaginary terms assure the 
                    preservation of boson commutation relations for 
                    the anihilation operators at the output of the 
                    beam splitter and that Eqs.~(2), (5a), etc.~of 
                    Ref.~\cite{om88} and Eqs.~(56), (57), etc.~of 
                    Ref.~\cite{ou88} should be corrected accordingly. 
                    Harry Paul also noticed the latter fact (private 
                    communication).}
\begin{eqnarray}
&&\hat E_3=\!\!\left(\hat a_{4x}\sqrt{T_x}\cos\theta_3+\hat
a_{4y}\sqrt{T_y}\sin\theta_3\right)\!
e^{i\,\omega_4({r_{_{II}}\!+r_3\over c}+t_{_0}^{^{II}}\!-t_3)}\nonumber\\
&&+i\left(\hat a_{3x}\sqrt{R_x}\cos\theta_3+\hat
a_{3y}\sqrt{R_y}\sin\theta_3\right)
e^{i\,\omega_3({r_{_{I}}+r_3\over c}+t_{_0}^{^{I}}-t_3)}\nonumber
\end{eqnarray}

With similar arguments the interactions leading to registration of a 
photon at detector D4 are given by the operator
\begin{eqnarray}
&&\hat E_4=\left(\hat a_{3x}\sqrt{T_x}\cos\theta_4+\hat
a_{3y}\sqrt{T_y}\sin\theta_4\right)
e^{i\,\omega_3({r_{_{I}}+r_4\over c}+t_{_0}^{^{I}}-t_4)}\nonumber\\
&&+i\left(\hat a_{4x}\sqrt{R_x}\cos\theta_4+\hat
a_{4y}\sqrt{R_y}\sin\theta_4\right)
e^{i\,\omega_4({r_{_{II}}+r_4\over c}+t_{_0}^{^{II}}-t_4)}\nonumber
\end{eqnarray}
Here, $r_I$ and $r_{II}$ denote the distance from the respective 
source to the beamsplitter, $r_3$ denotes the distance from the 
beamsplitter to detector D3, $t_3$ is the time of detection at 
D3, and $\omega_3$ is the frequency of photon 3. The symbols $r_4$, 
$t_4$ and $\omega_4$ are defined analogously. The evolution of the 
initial state through interaction with the whole setup including 
detection of one photon in each detector is then given by:
\begin{widetext}
\begin{eqnarray}
\hat E_4\hat E_3\hat E_2\hat E_1|\Psi\rangle
=&&e^{i\,[\omega_1({r_{_{1}}\over c}+t_{_0}^{^{I}}-t_1)+ 
\omega_2({r_{_{2}}\over c}+t_{_0}^{^{II}}-t_2)+ 
\omega_3({r_{_{I}}\over c}+t_{_0}^{^{I}})+ 
\omega_4({r_{_{II}}\over c}+t_{_0}^{^{II}})]}\nonumber\\
&\times&\bigl\{T_{14}T_{23}\,
e^{i\,[\omega_3({r_{_{4}}\over c}-t_{_{4}})+ 
\omega_4({r_{_{3}}\over c}-t_{_{3}})]}
-R_{24}R_{13}\,
e^{i\,[\omega_3({r_{_{3}}\over c}-t_{_{3}})+ 
\omega_4({r_{_{4}}\over c}-t_{_{4}})]}\bigr\}|0\rangle\label{eq:2}
\end{eqnarray}
\end{widetext}
where	
\begin{eqnarray}
T_{ij}=\sqrt{T_x}\cos\theta_i\cos\theta_j+ 
\sqrt{T_y}\sin\theta_i\sin\theta_j,\nonumber\\ 
R_{ij}=\sqrt{R_x}\cos\theta_i\cos\theta_j+ 
\sqrt{R_y}\sin\theta_i\sin\theta_j.
\end{eqnarray}

The squared modulus of this result gives the probability of having one 
photon arrive in each detector as a function of the angles of the 
polarizers. Note that the interference term, which is the real part of 
the product of the two terms in brackets, contains only the detection 
times $t_3$ and $t_4$ at detectors D3 and D4, respectively. 
Hence we would expect the 4th order interference to occur only in the 
coincidence counts of those two detectors. However, assuming for 
simplicity $r_I=r_{II}$, $r_3=r_4$, $\omega_3=\omega_4$, and 
$T_x=T_y=R_x=R_y=1/2$, we get for the coincidence probability:
\begin{eqnarray}
P(\theta_1,\theta_2,\theta_3,\theta_4)=&&
\langle\Psi|{\hat E}_{1}^{\dagger}{\hat E}_{2}^{\dagger}{\hat 
E}_3^{\dagger}{\hat E}_4^{\dagger}{\hat
E}_4{\hat E}_3{\hat E}_{2}{\hat E}_{1}|\Psi\rangle\nonumber\\
=&&{\textstyle{1\over16}}\sin^2(\theta_1-\theta_2)
\sin^2(\theta_3-\theta_4)\label{eq:3}
\end{eqnarray}

The correlations thus exist between the 
polarizations on the right side {\it and between those on the left 
side}. This means that the two photons going to the 
left can be forced into a nonlocal polarization correlation although 
they are emitted from two independent sources and nowhere share a 
common trajectory. Moreover, the correlation does not depend on the 
frequencies of the two photons, nor is there any condition as to the 
permissible time interval $|t_1-t_2|$ between their detections. 
This can be seen in Eq.~(\ref{eq:2}) where the relevant parameters 
$(\omega_1, \omega_2, t_1, t_2, r_1, r_2)$ only enter in the overall 
phase factor in contrast to photons 3 and 4, which must be 
detected within a time interval shorter than the beating period 
$|\omega_3-\omega_4|^{-1}$.

It is now interesting to see that the nonlocal polarization correlation 
between the photons on one side persists even when the polarizers on 
the other side are removed. Without the polarizers P1 and P2 of the 
left side, we have to sum over the probabilities of the four possible 
orthogonal settings of these polarizers and obtain
\begin{eqnarray}
P(\infty,\infty,\theta_3,\theta_4)
&=&{1\over8}\Bigl\{1-\cos^2(\theta_3-\theta_4)
\cos\bigl[(\omega_3-\omega_4)\nonumber\\
&\times&\bigl({r_4-r_3\over c}+
t_3-t_4\bigr)\bigr]\Bigr\}\label{eq:4}
\end{eqnarray}
where we set $r_I=r_{II}$ and $T_x=T_y=R_x=R_y=1/2$, but otherwise 
put no restrictions on the frequencies of the photons, the path 
lengths to the detectors, or the detection times. Eq.~(\ref{eq:4}) 
expresses the kind of nonlocal correlations from independent sources 
proposed by Yurke and Stoler \cite{yurke92}.

On the other hand, if we remove the polarizers of the right side, P3 and 
P4, we find 
\begin{eqnarray}
P(\theta_1,\theta_2,\infty,\infty)
&=&{1\over8}\Bigl\{1-\cos^2(\theta_1-\theta_2)
\cos\bigl[(\omega_3-\omega_4)\nonumber\\
&\times&\bigl({r_4-r_3\over c}+
t_3-t_4\bigr)\bigr]\Bigr\}\label{eq:5}
\end{eqnarray}
We see here explicitly, that the two photons on the left side can 
show a nonlocal correlation only if the time interval between 
detections on the right side, $|t_3-t_4|$, is kept below the beating 
frequency of the photons on the right side. This condition becomes 
trivial when we have $\omega_3=\omega_4$. Then the above expression 
suggests that there should be a time independent nonlocal polarization 
correlation between photons 1 and 2, although no polarization is 
measured on photons 3 and 4. At first glance this appears to be 
equivalent to having two independent 1-particle sources emitting 
unpolarized photons of different frequencies, which nowhere meet 
before their detection, and which yet are supposed to show nonlocal 
polarization correlations that are constant in time and we would 
expect no polarization correlation in such a situation. But also 
the result of Eq.~(\ref{eq:4}) is surprising, because it is what 
Ou and Mandel obtained \cite{om88,ohm88} for the interference of the 
two orthogonally polarized signal and idler beams of downconverted 
light, while we are dealing with originally unpolarized beams. 

In order to see whether the nonlocal correlations of 
eqs. (4) and (5) can be used to test Bell's inequality we turn to 
a more realistic description by means of wave packets. Each photon 
is represented by a gaussian amplitude distribution of energies. 
For the sake of simplicity all four photons shall have the same width 
of the energy distribution and thus the same coherence time T. 
The probability amplitude that photon $i$ has frequency $\omega_i$ 
when its central frequency is $\omega_{i0}$ is given by 
$$f(\omega_i,\omega_{i0},T)={\textstyle{{T^{1/2}\over\pi^{1/4}}}}
e^{-(\omega_i-\omega_{i0})^2T^2/2}\ \ \ \ \ \ \ \ \ \ 
{\rm with}\ i=1,\dots,4$$
where we normalized $|f_i|^2$ to 1. The final state of Eq.~(\ref{eq:2}) 
must be multiplied with these four functions and integrations must be 
made over the frequencies $\omega_i, i=1,...,4$. 
This models a photon pair as two wave packets fully overlapping at the 
source at the time of emission and then moving apart.
Now Eqs.~(4) and (5) turn into

\begin{eqnarray}
P(\infty,\infty,\theta_3,\theta_4)
&=&{F\over T^4}\Bigl\{\cosh(\tau_s\tau_{34}/T^2)\nonumber\\
&-&\cos^2(\theta_3-\theta_4)\cos\bigl[(\omega_3-\omega_4)
\tau_{34}\bigr]\Bigr\}\qquad\label{eq:4'}\\
P(\theta_1,\theta_2,\infty,\infty)
&=&{F\over T^4}\Bigl\{\cosh(\tau_s\tau_{34}/T^2)\nonumber\\
&-&\cos^2(\theta_1-\theta_2)\cos\bigl[(\omega_3-\omega_4)
\tau_{34}\bigr]\Bigr\}\qquad\label{eq:5'}
\end{eqnarray}
where we set $r_3=r_4$ and defined $\tau_s \equiv t_0^I - t_0^{II}$ 
and $\tau_{34} \equiv t_3-t_4$. 
We are now dealing with probability {\it densities} for a quadruple 
detection at time {\it points} $t_1$, $t_2$, $t_3$ and $t_4$. 
The damping term $F$ contains the detection times $t_1$ and $t_2$ 
and other experimental parameters. It expresses how well the wavepackets 
are centered at the various detectors at the respective detection times. 
Note that in both cases the polarization correlations persist but are 
reduced in visibility, which is given by
$v=\bigl[2\cosh(\tau_s\tau_{34}/T^2)-1\bigr]^{-1}$ 
when $(\omega_{30}-\omega_{40})\tau_{34}=0$. For a violation of Bell's 
inequality, and hence for a possible exclusion of hidden variable 
theories, $v$ must be larger than $2^{-1/2}$ implying the product 
$\tau_{34}\tau_s$ must be less than $0.663 T^2$. However, the time 
interval between the emissions at the two sources, $\tau_s$, is not a 
directly measurable quantity, but must be inferred from the detection 
times. Therefore $\tau_s$ cannot be known better than to about $T$. 
Hence $\tau_{34}<<T$ is a necessary requirement, 
which means D3 and D4 must fire in ultra short coincidence. 
In consequence most of the data collected at D3 and D4 must be 
discarded, as even for simultaneous emissions from the sources the 
mean value of $\tau_{34}$ is about $\sqrt{2}T$. This in turn 
precludes a test of local hidden variable theories by means of 
Bell's inequality at detectors D3 and D4, where a postselection 
throws away more than 31\%\ of the data \cite{gam87}. 
On the other hand, the test is possible at detectors D1 and 
D2 [Eq.~(\ref{eq:5'})]. Again the requirement is for ultra short 
coincidence at detectors D3 and D4 and not at D1 and D2, where there 
is no upper bound on the coincidence window $|t_1 - t_2|$. In fact 
one would permit a wide coincidence window in order to collect all 
data at D1 and D2 that have been preselected by the ultra short 
coincidence between D3 and D4. This constitutes a most 
discriminating test of Bell's inequality, because no postselection 
of the nonlocally correlated particles 1 and 2 is needed.

Let us now conclude with an attempt to understand how the polarization 
correlation of the particles on the left side can come about even when 
there are no polarizers on the right side and why only the 
mutual angles on each side --- Eq.~(\ref{eq:3}) --- are relevant for 
the overall coincidences. The answer lies in the 
beam splitter. It superimposes the states of photon 3 and 4. The beams 
going to detectors D3 and D4 must therefore reflect the bosonic character 
of the particles and can only be occupied in the following two ways:

(a) \it Both photons in the same beam.\/ \rm 
The commutation rules demand that the two photons, e.g., from the beam 
going to D3 (assuming $r_I=r_{II}$, $r_3=r_4$, and $T_x=T_y=R_x=R_y=1/2$) 
obey the following counterpart of Eq.~(\ref{eq:3}): 
\begin{eqnarray}
P(\theta_1,\theta_2,2\times\theta_3)=&&
\langle\Psi|{\hat E}_{1}^{\dagger}{\hat E}_{2}^{\dagger}{\hat 
E}_3^{\dagger}{\hat E}_3^{\dagger}{\hat
E}_3{\hat E}_3{\hat E}_{2}{\hat E}_{1}|\Psi\rangle\nonumber\\
={\textstyle{1\over4}}\cos^2(\theta_1&&-\theta_3)\cos^2
(\theta_2-\theta_3)
\end{eqnarray}
 
given they have the same energy ($\omega_3=\omega_4$). 
When there are no polarizers before D3 and D4, photons 1 and 2 going to 
the left would still have the polarization correlation, but 
individually they would be unpolarized. For both 2--photons channels  
together we obtain $P(\theta_1,\theta_2,\infty,\infty)=
[1+\cos^2(\theta_1-\theta_2)]/8$. 

(b)\it One photon in each beam: 1-1--channel.\/ \rm Now the 
commutation rules require that the polarizations are correlated as 
given by Eq.~(\ref{eq:3}). This means we also get the correlation 
between the polarizations of beams 1 and 2 as given by Eq.~(\ref{eq:5}):   
$P(\theta_1,\theta_2,\infty,\infty)=
\sin^2(\theta_1-\theta_2)/8$, i.e., with the maximum in the
orthogonal direction with regard to the previous case.

Of course, the polarization correlation between photons 1 and 2 
disappears if we do not detect them in coincidence detections on the 
right side, which one can see from the fact that the sum of 
$P(\theta_1,\theta_2,\infty,\infty)$ from all the channels is a constant.  

To understand Eq.~(\ref{eq:3}) let us compare it with the standard 
\it left--right\/ \rm Bell probabilities:  
$P(\theta_1,\theta_3)={1\over2}cos^2(\theta_1-\theta_3)$ 
and $P(\theta_2,\theta_4)={1\over2}cos^2(\theta_2-\theta_4)$.   
For the angles $\theta_1=\theta_2$ and 
$\theta_3=\theta_4$ we obtain [from Eq.~(\ref{eq:3})]  
$P(\theta_1,\theta_3,\theta_2,\theta_4)=0$ no matter the values of 
$\theta_1-\theta_3$ and $\theta_2-\theta_4$.  
These three probabilities clearly cannot be satisfied simultaneously 
and we can express this fact by saying that the 4th order interference 
\it erases\/ \rm information on polarization correlation in the 
\it 1-1--channel\/\rm. An immediate consequence is that for 
$\theta_1=\theta_2=\theta_3=\theta_4$ we can never register
coincidences and this represents yet another possibility to formulate 
Bell's theorem without inequalites --- the idea first developed  
by Greenberger, Horne, and Zeilinger \cite{ghz87}. The possiblity 
is based on the well--known fact that the classical visibility of 
the 4th order interference is not higher than 50\%.

It is interesting that although for  $\omega_3\ne\omega_4$ we can track 
down the Bell \it left--right\/ \rm probabilities in the 2--photons
beams exactly, as expected, in the \it 1-1--channel\/ \rm this is even 
then not possible. The case of different energies can be handled
with frequency filters (FF in Fig.~1). The result is then modulated 
with the beating period but in effect we have 
$P(\theta_1,\theta_2,\theta_3,\theta_4)={\textstyle{1\over2}}\cos^2
(\theta_1-\theta_3)\cos^2(\theta_2-\theta_4)$ if we assume two 
polarizers and two detectors in each 2--photons beam (not shown in 
Fig.~1). So, dropped polarizer in front of the one of the four detectors 
immediately gives the standard \it left--right\/ \rm Bell probability for 
the other pair. The \it 1-1--channel\/\rm, on the other hand, responds to 
the special feature of the interference of the 4th order to ``create'' the 
polarization correlation even when unpolarized photons interfere --- 
see Eq.~(\ref{eq:4}). 

Thus, while the 2nd order interference erases the \it path\/ \rm 
memory, the 4th order interference erases the \it 
polarization correlation\/ \rm memory.
It occurs in an analogous way in which the 4th order interference
erases the \it polarization\/ \rm memory of two polarized incident
photons according to Eq.~(16) of Ref.~\cite{p-pra94}.

One of us (M.P.) is grateful to his host K.-E.~Hellwig,
Inst.~Theor.~Physics, TU Berlin where he completed 
the first draft of the present paper containing the full 
elaboration in the plane waves and discussed it in a series of 4 
seminars he held at TU Berlin (Jul 16 - Aug 9, 1993). 

%


\begin{thebibliography}{22}%
\makeatletter
\providecommand \@ifxundefined [1]{%
 \@ifx{#1\undefined}
}%
\providecommand \@ifnum [1]{%
 \ifnum #1\expandafter \@firstoftwo
 \else \expandafter \@secondoftwo
 \fi
}%
\providecommand \@ifx [1]{%
 \ifx #1\expandafter \@firstoftwo
 \else \expandafter \@secondoftwo
 \fi
}%
\providecommand \natexlab [1]{#1}%
\providecommand \enquote  [1]{``#1''}%
\providecommand \bibnamefont  [1]{#1}%
\providecommand \bibfnamefont [1]{#1}%
\providecommand \citenamefont [1]{#1}%
\providecommand \href@noop [0]{\@secondoftwo}%
\providecommand \href [0]{\begingroup \@sanitize@url \@href}%
\providecommand \@href[1]{\@@startlink{#1}\@@href}%
\providecommand \@@href[1]{\endgroup#1\@@endlink}%
\providecommand \@sanitize@url [0]{\catcode `\\12\catcode `\$12\catcode
  `\&12\catcode `\#12\catcode `\^12\catcode `\_12\catcode `\%12\relax}%
\providecommand \@@startlink[1]{}%
\providecommand \@@endlink[0]{}%
\providecommand \url  [0]{\begingroup\@sanitize@url \@url }%
\providecommand \@url [1]{\endgroup\@href {#1}{\urlprefix }}%
\providecommand \urlprefix  [0]{URL }%
\providecommand \Eprint [0]{\href }%
\@ifxundefined \urlstyle {%
  \providecommand \doi  [0]{\begingroup \@sanitize@url \@doi}%
  \providecommand \@doi [1]{\endgroup \@@startlink {\doibase
  #1}doi:\discretionary {}{}{}#1\@@endlink }%
}{%
  \providecommand \doi  [0]{doi:\discretionary{}{}{}\begingroup
  \urlstyle{rm}\Url }%
}%
\providecommand \doibase [0]{http://dx.doi.org/}%
\providecommand \Doi [0]{\begingroup \@sanitize@url \@Doi }%
\providecommand \@Doi  [1]{\endgroup\@@startlink{\doibase#1}\@@Doi}%
\providecommand \@@Doi [1]{#1\@@endlink}%
\providecommand \selectlanguage [0]{\@gobble}%
\providecommand \bibinfo  [0]{\@secondoftwo}%
\providecommand \bibfield  [0]{\@secondoftwo}%
\providecommand \translation [1]{[#1]}%
\providecommand \BibitemOpen [0]{}%
\providecommand \bibitemStop [0]{}%
\providecommand \bibitemNoStop [0]{.\EOS\space}%
\providecommand \EOS [0]{\spacefactor3000\relax}%
\providecommand \BibitemShut  [1]{\csname bibitem#1\endcsname}%
\bibitem [{\citenamefont {Ou}(1988)}]{ou88}%
  \BibitemOpen
  \bibfield  {author} {\bibinfo {author} {\bibfnamefont {Z.~Y.}\ \bibnamefont
  {Ou}},\ }\href@noop {} {\bibfield  {journal} {\bibinfo  {journal} {{\it Phys.
  Rev. A}},\ }\textbf {\bibinfo {volume} {{\bf 37}}},\ \bibinfo {pages} {1607}
  (\bibinfo {year} {1988})}\BibitemShut {NoStop}%
\bibitem [{\citenamefont {Mandel}(1983)}]{man83}%
  \BibitemOpen
  \bibfield  {author} {\bibinfo {author} {\bibfnamefont {L.}~\bibnamefont
  {Mandel}},\ }\href@noop {} {\bibfield  {journal} {\bibinfo  {journal} {{\it
  Phys. Rev. A}},\ }\textbf {\bibinfo {volume} {{\bf 28}}},\ \bibinfo {pages}
  {929} (\bibinfo {year} {1983})}\BibitemShut {NoStop}%
\bibitem [{\citenamefont {Paul}(1986)}]{paul86}%
  \BibitemOpen
  \bibfield  {author} {\bibinfo {author} {\bibfnamefont {H.}~\bibnamefont
  {Paul}},\ }\href@noop {} {\bibfield  {journal} {\bibinfo  {journal} {{\it
  Rev. Mod. Phys.}},\ }\textbf {\bibinfo {volume} {{\bf 58}}},\ \bibinfo
  {pages} {209} (\bibinfo {year} {1986})}\BibitemShut {NoStop}%
\bibitem [{\citenamefont {Rarity}\ \emph {et~al.}(1990)\citenamefont {Rarity},
  \citenamefont {Tapster}, \citenamefont {Larchuk}, \citenamefont {Campos},
  \citenamefont {Teich},\ and\ \citenamefont {Saleh}}]{rari90}%
  \BibitemOpen
  \bibfield  {author} {\bibinfo {author} {\bibfnamefont {J.~G.}\ \bibnamefont
  {Rarity}}, \bibinfo {author} {\bibfnamefont {P.~R.}\ \bibnamefont {Tapster}},
  \bibinfo {author} {\bibfnamefont {E.~J.~T.}\ \bibnamefont {Larchuk}},
  \bibinfo {author} {\bibfnamefont {R.~A.}\ \bibnamefont {Campos}}, \bibinfo
  {author} {\bibfnamefont {M.~C.}\ \bibnamefont {Teich}}, \ and\ \bibinfo
  {author} {\bibfnamefont {B.~E.~A.}\ \bibnamefont {Saleh}},\ }\href@noop {}
  {\bibfield  {journal} {\bibinfo  {journal} {{\it Phys. Rev. Lett.}},\
  }\textbf {\bibinfo {volume} {{\bf 65}}},\ \bibinfo {pages} {1348–}
  (\bibinfo {year} {1990})}\BibitemShut {NoStop}%
\bibitem [{\citenamefont {Ou}\ \emph {et~al.}(1990){\natexlab{a}}\citenamefont
  {Ou}, \citenamefont {Zou}, \citenamefont {Wang},\ and\ \citenamefont
  {Mandel}}]{ozwm90}%
  \BibitemOpen
  \bibfield  {author} {\bibinfo {author} {\bibfnamefont {Z.~Y.}\ \bibnamefont
  {Ou}}, \bibinfo {author} {\bibfnamefont {X.~Y.}\ \bibnamefont {Zou}},
  \bibinfo {author} {\bibfnamefont {L.~J.}\ \bibnamefont {Wang}}, \ and\
  \bibinfo {author} {\bibfnamefont {L.}~\bibnamefont {Mandel}},\ }\href@noop {}
  {\bibfield  {journal} {\bibinfo  {journal} {{\it Phys. Rev. A}},\ }\textbf
  {\bibinfo {volume} {{\bf 42}}},\ \bibinfo {pages} {2957} (\bibinfo {year}
  {1990}{\natexlab{a}})}\BibitemShut {NoStop}%
\bibitem [{\citenamefont {Ou}\ \emph {et~al.}(1990){\natexlab{b}}\citenamefont
  {Ou}, \citenamefont {Wang}, \citenamefont {Zou},\ and\ \citenamefont
  {Mandel}}]{owzm90}%
  \BibitemOpen
  \bibfield  {author} {\bibinfo {author} {\bibfnamefont {Z.~Y.}\ \bibnamefont
  {Ou}}, \bibinfo {author} {\bibfnamefont {L.~J.}\ \bibnamefont {Wang}},
  \bibinfo {author} {\bibfnamefont {X.~Y.}\ \bibnamefont {Zou}}, \ and\
  \bibinfo {author} {\bibfnamefont {L.}~\bibnamefont {Mandel}},\ }\href@noop {}
  {\bibfield  {journal} {\bibinfo  {journal} {{\it Phys. Rev. A}},\ }\textbf
  {\bibinfo {volume} {{\bf 41}}},\ \bibinfo {pages} {566} (\bibinfo {year}
  {1990}{\natexlab{b}})}\BibitemShut {NoStop}%
\bibitem [{\citenamefont {Ou}\ and\ \citenamefont {Mandel}(1988)}]{om88}%
  \BibitemOpen
  \bibfield  {author} {\bibinfo {author} {\bibfnamefont {Z.~Y.}\ \bibnamefont
  {Ou}}\ and\ \bibinfo {author} {\bibfnamefont {L.}~\bibnamefont {Mandel}},\
  }\href@noop {} {\bibfield  {journal} {\bibinfo  {journal} {{\it Phys. Rev.
  Lett.}},\ }\textbf {\bibinfo {volume} {{\bf 61}}},\ \bibinfo {pages} {50}
  (\bibinfo {year} {1988})}\BibitemShut {NoStop}%
\bibitem [{\citenamefont {Ou}\ \emph {et~al.}(1988)\citenamefont {Ou},
  \citenamefont {Hong},\ and\ \citenamefont {Mandel}}]{ohm88}%
  \BibitemOpen
  \bibfield  {author} {\bibinfo {author} {\bibfnamefont {Z.~Y.}\ \bibnamefont
  {Ou}}, \bibinfo {author} {\bibfnamefont {C.}~\bibnamefont {Hong}}, \ and\
  \bibinfo {author} {\bibfnamefont {L.}~\bibnamefont {Mandel}},\ }\href@noop {}
  {\bibfield  {journal} {\bibinfo  {journal} {{\it Opt. Commun.}},\ }\textbf
  {\bibinfo {volume} {{\bf 67}}},\ \bibinfo {pages} {169} (\bibinfo {year}
  {1988})}\BibitemShut {NoStop}%
\bibitem [{\citenamefont {Ou}\ \emph {et~al.}(1987)\citenamefont {Ou},
  \citenamefont {Hong},\ and\ \citenamefont {Mandel}}]{ohm87}%
  \BibitemOpen
  \bibfield  {author} {\bibinfo {author} {\bibfnamefont {Z.~Y.}\ \bibnamefont
  {Ou}}, \bibinfo {author} {\bibfnamefont {C.}~\bibnamefont {Hong}}, \ and\
  \bibinfo {author} {\bibfnamefont {L.}~\bibnamefont {Mandel}},\ }\href@noop {}
  {\bibfield  {journal} {\bibinfo  {journal} {{\it Phys. Lett. A}},\ }\textbf
  {\bibinfo {volume} {{\bf 122}}},\ \bibinfo {pages} {11} (\bibinfo {year}
  {1987})}\BibitemShut {NoStop}%
\bibitem [{\citenamefont {Campos}\ \emph {et~al.}(1990)\citenamefont {Campos},
  \citenamefont {Saleh},\ and\ \citenamefont {Teich}}]{cst90}%
  \BibitemOpen
  \bibfield  {author} {\bibinfo {author} {\bibfnamefont {R.~A.}\ \bibnamefont
  {Campos}}, \bibinfo {author} {\bibfnamefont {B.~E.~A.}\ \bibnamefont
  {Saleh}}, \ and\ \bibinfo {author} {\bibfnamefont {M.~C.}\ \bibnamefont
  {Teich}},\ }\href@noop {} {\bibfield  {journal} {\bibinfo  {journal} {{\it
  Phys. Rev. A}},\ }\textbf {\bibinfo {volume} {{\bf 42}}},\ \bibinfo {pages}
  {4127–} (\bibinfo {year} {1990})}\BibitemShut {NoStop}%
\bibitem [{\citenamefont {Horne}\ \emph {et~al.}(1989)\citenamefont {Horne},
  \citenamefont {Shimony},\ and\ \citenamefont {Zeilinger}}]{hsz89}%
  \BibitemOpen
  \bibfield  {author} {\bibinfo {author} {\bibfnamefont {M.~A.}\ \bibnamefont
  {Horne}}, \bibinfo {author} {\bibfnamefont {A.}~\bibnamefont {Shimony}}, \
  and\ \bibinfo {author} {\bibfnamefont {A.}~\bibnamefont {Zeilinger}},\
  }\href@noop {} {\bibfield  {journal} {\bibinfo  {journal} {{\it Phys. Rev.
  Lett}},\ }\textbf {\bibinfo {volume} {{\bf 62}}},\ \bibinfo {pages} {2209–}
  (\bibinfo {year} {1989})}\BibitemShut {NoStop}%
\bibitem [{\citenamefont {Wang}\ \emph
  {et~al.}(1991){\natexlab{a}}\citenamefont {Wang}, \citenamefont {Zou},\ and\
  \citenamefont {Mandel}}]{wa91}%
  \BibitemOpen
  \bibfield  {author} {\bibinfo {author} {\bibfnamefont {L.~J.}\ \bibnamefont
  {Wang}}, \bibinfo {author} {\bibfnamefont {X.~Y.}\ \bibnamefont {Zou}}, \
  and\ \bibinfo {author} {\bibfnamefont {L.}~\bibnamefont {Mandel}},\
  }\href@noop {} {\bibfield  {journal} {\bibinfo  {journal} {{\it Phys. Rev.
  Lett.}},\ }\textbf {\bibinfo {volume} {{\bf 66}}},\ \bibinfo {pages} {1111}
  (\bibinfo {year} {1991}{\natexlab{a}})}\BibitemShut {NoStop}%
\bibitem [{\citenamefont {Holland}\ and\ \citenamefont {Vigier}(1991)}]{hv91}%
  \BibitemOpen
  \bibfield  {author} {\bibinfo {author} {\bibfnamefont {P.~R.}\ \bibnamefont
  {Holland}}\ and\ \bibinfo {author} {\bibfnamefont {J.~P.}\ \bibnamefont
  {Vigier}},\ }\href@noop {} {\bibfield  {journal} {\bibinfo  {journal} {{\it
  Phys. Rev. Lett.}},\ }\textbf {\bibinfo {volume} {{\bf 67}}},\ \bibinfo
  {pages} {402} (\bibinfo {year} {1991})}\BibitemShut {NoStop}%
\bibitem [{\citenamefont {Wang}\ \emph
  {et~al.}(1991){\natexlab{b}}\citenamefont {Wang}, \citenamefont {Zou},\ and\
  \citenamefont {Mandel}}]{wzm91}%
  \BibitemOpen
  \bibfield  {author} {\bibinfo {author} {\bibfnamefont {L.~J.}\ \bibnamefont
  {Wang}}, \bibinfo {author} {\bibfnamefont {X.~Y.}\ \bibnamefont {Zou}}, \
  and\ \bibinfo {author} {\bibfnamefont {L.}~\bibnamefont {Mandel}},\
  }\href@noop {} {\bibfield  {journal} {\bibinfo  {journal} {{\it Phys. Rev.
  Lett.}},\ }\textbf {\bibinfo {volume} {{\bf 67}}},\ \bibinfo {pages} {403}
  (\bibinfo {year} {1991}{\natexlab{b}})}\BibitemShut {NoStop}%
\bibitem [{\citenamefont {Silverman}(1993)}]{silv93}%
  \BibitemOpen
  \bibfield  {author} {\bibinfo {author} {\bibfnamefont {M.~P.}\ \bibnamefont
  {Silverman}},\ }\href@noop {} {\bibfield  {journal} {\bibinfo  {journal}
  {{\it Am. J. Phys.}},\ }\textbf {\bibinfo {volume} {{\bf 61}}},\ \bibinfo
  {pages} {514} (\bibinfo {year} {1993})}\BibitemShut {NoStop}%
\bibitem [{\citenamefont {Bell}(1986)}]{bell86}%
  \BibitemOpen
  \bibfield  {author} {\bibinfo {author} {\bibfnamefont {J.~S.}\ \bibnamefont
  {Bell}},\ }\href@noop {} {\bibfield  {journal} {\bibinfo  {journal} {{\it
  Phys. Reports}},\ }\textbf {\bibinfo {volume} {{\bf 137}}},\ \bibinfo {pages}
  {7} (\bibinfo {year} {1986})}\BibitemShut {NoStop}%
\bibitem [{\citenamefont {Pavi{\v c}i{\'c}}(1994)}]{p-pra94}%
  \BibitemOpen
  \bibfield  {author} {\bibinfo {author} {\bibfnamefont {M.}~\bibnamefont
  {Pavi{\v c}i{\'c}}},\ }\href@noop {} {\bibfield  {journal} {\bibinfo
  {journal} {{\it Phys. Rev. A}},\ }\textbf {\bibinfo {volume} {{\bf 50}}},\
  \bibinfo {pages} {3486} (\bibinfo {year} {1994})}\BibitemShut {NoStop}%
\bibitem [{\citenamefont {Aspect}\ \emph {et~al.}(1982)\citenamefont {Aspect},
  \citenamefont {Dalibard},\ and\ \citenamefont {Roger}}]{aspect2}%
  \BibitemOpen
  \bibfield  {author} {\bibinfo {author} {\bibfnamefont {A.}~\bibnamefont
  {Aspect}}, \bibinfo {author} {\bibfnamefont {J.}~\bibnamefont {Dalibard}}, \
  and\ \bibinfo {author} {\bibfnamefont {G.}~\bibnamefont {Roger}},\
  }\href@noop {} {\bibfield  {journal} {\bibinfo  {journal} {{\it Phys. Rev.
  Lett.}},\ }\textbf {\bibinfo {volume} {{\bf 49}}},\ \bibinfo {pages} {1804}
  (\bibinfo {year} {1982})}\BibitemShut {NoStop}%
\bibitem [{Note1()}]{Note1}%
  \BibitemOpen
  \bibinfo {note} {For the given operators cf.~Refs.~\protect \cite
  {ou88,owzm90,om88,ohm88} Note that positive imaginary terms assure the
  preservation of boson commutation relations for the anihilation operators at
  the output of the beam splitter and that Eqs.~(2), (5a), etc.~of Ref.~\cite
  {om88} and Eqs.~(56), (57), etc.~of Ref.~\cite {ou88} should be corrected
  accordingly. Harry Paul also noticed the latter fact (private
  communication).}\BibitemShut {Stop}%
\bibitem [{\citenamefont {Yurke}\ and\ \citenamefont {Stoler}(1992)}]{yurke92}%
  \BibitemOpen
  \bibfield  {author} {\bibinfo {author} {\bibfnamefont {B.}~\bibnamefont
  {Yurke}}\ and\ \bibinfo {author} {\bibfnamefont {D.}~\bibnamefont {Stoler}},\
  }\href@noop {} {\bibfield  {journal} {\bibinfo  {journal} {{\it Phys. Rev.
  Lett.}},\ }\textbf {\bibinfo {volume} {{\bf 68}}},\ \bibinfo {pages}
  {1251–} (\bibinfo {year} {1992})}\BibitemShut {NoStop}%
\bibitem [{\citenamefont {Garg}\ and\ \citenamefont {Mermin}(1987)}]{gam87}%
  \BibitemOpen
  \bibfield  {author} {\bibinfo {author} {\bibfnamefont {A.}~\bibnamefont
  {Garg}}\ and\ \bibinfo {author} {\bibfnamefont {N.~D.}\ \bibnamefont
  {Mermin}},\ }\href@noop {} {\bibfield  {journal} {\bibinfo  {journal} {{\it
  Phys. Rev. D}},\ }\textbf {\bibinfo {volume} {{\bf 35}}},\ \bibinfo {pages}
  {3831–} (\bibinfo {year} {1987})}\BibitemShut {NoStop}%
\bibitem [{\citenamefont {Greenberger}\ \emph {et~al.}(1989)\citenamefont
  {Greenberger}, \citenamefont {Horne},\ and\ \citenamefont
  {Zeilinger}}]{ghz87}%
  \BibitemOpen
  \bibfield  {author} {\bibinfo {author} {\bibfnamefont {D.~M.}\ \bibnamefont
  {Greenberger}}, \bibinfo {author} {\bibfnamefont {M.}~\bibnamefont {Horne}},
  \ and\ \bibinfo {author} {\bibfnamefont {A.}~\bibnamefont {Zeilinger}},\ }in\
  \href@noop {} {\emph {\bibinfo {booktitle} {Bell's Theorem, Quantum Theory,
  and Conceptions of the Universe}}},\ \bibinfo {editor} {edited by\ \bibinfo
  {editor} {\bibfnamefont {N.}~\bibnamefont {Kafatos}}}\ (\bibinfo  {publisher}
  {Kluwer},\ \bibinfo {address} {Dordrect},\ \bibinfo {year} {1989})\ pp.\
  \bibinfo {pages} {69--72}\BibitemShut {NoStop}%
\end{thebibliography}

\end{document}